\documentclass[a4paper,11pt]{article}
\usepackage[utf8]{inputenc}
\usepackage{cite}
\usepackage{wrapfig}
\usepackage{a4wide}
\usepackage{amssymb}
\usepackage{amsmath,dsfont}
\usepackage{mathrsfs}
\RequirePackage[colorlinks=true,urlcolor=blue,anchorcolor=blue,citecolor=blue,filecolor=blue,linkcolor=blue,menucolor=blue,pdfproducer=medialab,pdfa=true]{hyperref}
\usepackage{pdflscape}
\usepackage{tikz}
\usetikzlibrary{positioning}
\usetikzlibrary{arrows}
\usepackage{array} 
\usepackage{multirow} 
\usepackage{colortbl} 
\usepackage{arydshln} 
\usepackage{rotating} 
\usepackage{comment}
\usetikzlibrary{arrows.meta}

\tikzset{miniG/.style={draw=none,minimum size=1pt,fill=black,circle,inner sep=3pt, draw}}
\usepackage{oubraces}
\usetikzlibrary{shapes.geometric}
\usetikzlibrary{shapes.misc}
\tikzset{gauge1/.style={draw=none,minimum size=0.4cm,fill=white,circle, draw}}
\tikzset{flavour1/.style={draw=none,minimum size=0.5cm,fill=white, regular polygon,regular polygon sides=4,draw}}
\tikzset{none/.style={draw=none}}
\pgfdeclarelayer{edgelayer}
\pgfdeclarelayer{nodelayer}
\usetikzlibrary{decorations.pathreplacing}
\pgfsetlayers{edgelayer,nodelayer,main}
\def\blue#1{{\color{blue}{#1}}}
\def\green#1{{\color{black!25!green}{#1}}}

\def\rcy{\rowcolor{black!25!yellow!10}}

\def\rcr{\rowcolor{red!07}}

\def\bar{\overline}

\def\^{\wedge}
\def\I{\mathds{1}}

\def\U{{\rm U}}

\def\Sp{\mathop{\rm Sp}}

\def\C{\mathbb{C}} 

\def\N{\mathbb{N}}

\def\R{\mathbb{R}} 
 
\def\Z{\mathbb{Z}}

\def\ff{{\mathfrak f}}

\def\cC{{\mathcal C}}

\def\cM{{\mathcal M}}
\def\cN{{\mathcal N}}

\def\cV{{\mathcal V}}

\def\a{{\alpha}}

\def\g{{\gamma}}

\def\D{{\Delta}}

\def\s{{\sigma}}

\def\vf{{\varphi}}

\def\beq{\begin{equation}}
\def\eeq{\end{equation}}

\def\I{\mathds{1}}

\begin{document}
\begin{center}
\huge 
\textbf{Construction and classification of }\\ \vspace{0.5em}
\textbf{Coulomb branch geometries}
\normalsize

\vspace{1em}
Philip C. Argyres and Mario Martone
 
\vspace{1em}
\small
\textit{University of Cincinnati, Physics Department, Cincinnati OH 45221\\\vspace{.4em}
University of Texas, Austin, Physics Department, Austin TX 78712.\\\vspace{.4em}}
\normalsize
\end{center}

\begin{abstract}
We give a non-technical summary of the classification program, very dear to the hearts of both authors, of four dimensional $\cN=2$ superconformal field theories (SCFTs) based on the study of their Coulomb branch geometries. We outline the main ideas behind this program, review the most important results thus far obtained \cite{
Argyres:2015ffa,
Argyres:2015gha,
Argyres:2016xua,
Argyres:2016xmc,
Argyres:2016yzz,
Argyres:2017tmj,
Argyres:2018zay,
Caorsi:2018zsq,
Argyres:2018urp,
Caorsi:2018ahl,
Bourget:2019phe,
Argyres:2018wxu,
Argyres:2019ngz,
Caorsi:2019vex,
Argyres:2019yyb}, and the prospects for future results. This contribution will appear in the volume \emph{the Pollica perspective on the (super)-conformal world} but we decided to also make it available separately in the hope that it could be useful to those who are interested in obtaining a quick grasp of this rapidly developing program.
\end{abstract}

\section{Logic of the program}

The complex scalars in $\cN=2$ gauge theories in four dimensions generically admit continuous vacuum expectation values which are not lifted by quantum corrections. This space of vacuum solutions is called the moduli spaces of vacua, $\cM$. Depending on the properties of the low-energy physics, $\cM$ is divided into \emph{Coulomb}, \emph{Higgs} and \emph{mixed} branches. We focus on the former and indicate the Coulomb branch (CB) of an $\cN=2$ theory by $\cC$. The properties of both $\cM$ and $\cC$ can be formulated in a way that makes no reference to the quantum fields in the gauge theory. Therefore characterizing $\cN=2$ field theories via the properties of their moduli space is immediately suitable to study theories with no known lagrangian formulation as is the case for the majority of $\cN=2$ superconformal field theories (SCFTs) which are the ultimate objective of our study. 

This note is organized as follows. In this first section, we provide a lighting review of the CB and outline in some detail the logic of the classification program. In the second section, we report some of the most important results which our analysis has thus far produced. We conclude by outlining the directions which we are currently pursuing and identify the main open questions of this program.

\paragraph{Coulomb Branch generalities.}

The property defining $\cC$ is that the low-energy theory on a generic point is extremely simple: it is just a free $\cN=2$ supersymmetric $U(1)^r$ gauge theory with no massless charged states. $r$ is called the \emph{rank} of the theory and coincides with the complex dimensionality of $\cC$, ${\rm dim}_\C\cC=r$. $\cC$ is a singular space and its (complex co-dimension one) singular locus, $\cV$, is the locus where charged states become massless. In other words, $\cV$ represents precisely the locus where the low-energy physics is less boring and even potentially interesting: it may no longer be free.  Interacting scale-invariant physics is often hard to characterize directly, as it typically does not have a useful lagrangian description.  But the geometry of $\cC$, though singular at this locus, is amenable to analysis, and so provides some information about these interacting scale-invariant theories. 

The striking fact about CB geometry is that the physics on $\cV$ can be studied in a fairly detailed way by studying the theory in the non-singular region, $\cC_{\rm reg}:=\cC\setminus \cV$ where the low-energy physics is as simple as it gets (just a bunch of non-interacting $\cN=2$ vector multiplets!). This is due to the fact that no globally defined lagrangian description of the low energy $\cN=2$ $U(1)^r$ is possible and non-trivial \emph{monodromies} have to considered to describe the physics on $\cC_{\rm reg}$. These are specific elements of the $\Sp(2r,\Z)$ \emph{electromagnetic duality group} which depend on the physics at $\cV$ and can be therefore used to characterize it. The object which transforms non-trivially under the monodromy group is the vector of \emph{special coordinates} $\s$, which provides a holomorphic section of an $\Sp(2r,\Z)$ bundle over $\cC_{\rm reg}$. The special coordinates also satisfy non trivial constraints which allow the definition of a K\"ahler metric on $\cC_{\rm reg}$ and which can be extended in a non-trivial way to $\cV$ (see the {\bf stratification} section below). All this together equips $\cC_{\rm reg}$ with a \emph{rigid special K\"ahler} (RSK) structure.

$\cN=2$ SCFTs live at the origin of scale invariant CBs. Scale invariance makes the study of the geometry in the non-singular locus even simpler and strongly constrains the allowed monodromies. Our program aims at extracting the most information with the minimum (which for ranks greater than one is still substantial) effort, and characterize the space of $\cN=2$ SCFTs by understanding the properties of the non-singular region of their CB geometries.


Other facts further motivate our approach. There is a belief that all interacting $\cN=2$ SCFT have a CB (see the {\bf rank-0} section below), and thus can be captured by our classification method. This should be contrasted with the other branches of the moduli space, where an infinite number of interacting $\cN=2$ SCFTs with trivial Higgs and/or mixed branches are known.  Also, interestingly, the CB has the property that it is only deformed and not lifted by $\cN=2$-preserving relevant deformations. Thus the RG-structure of $\cN=2$ theories is immediately visible from the CB geometry. 

There is a natural way to organize our classification program. First, theories with lower-dimensional CBs are simpler, and in particular lower-rank theories can be reached via RG flows of higher ones but not vice versa (see the {\bf stratification} section below).  Thus it is reasonable to study CB geometries in order of their increasing dimensionality.\footnote{The story is different for the Higgs branch.  There the natural organizing parameter is the number of symplectic leaves rather than the overall Higgs branch dimension.} 

Secondly, if conformal invariance is unbroken, scale invariance of the corresponding geometry dramatically constrains its global structure. Quick progress can be made by classifying the \emph{scale invariant limit} of these geometries. In fact for $\cN=2$ SCFTs, the $\R^+\times \U(1)$ action by dilatations and the $U(1)_R$ symmetry induces a $\C^*$ action on the full CB geometry. In the one-dimensional case, requiring invariance under this $\C^*$ action immediately constrains the set of allowed geometries to be one of only 7 possibilities. We discuss the two-dimensional case below.

Unfortunately it is a fact, known since the seminal papers on the topic \cite{Seiberg:1994rs,Seiberg:1994aj}, that many SCFTs share the same scale invariant CB geometry, and therefore this information is not enough to fully characterize the space of $\cN=2$ SCFTs.  Also in the conformal limit, $\cC_{\rm reg}$ is ``so simple'' that not much information on the non-trivial physics of the $\cN=2$ SCFT living at the origin of $\cC$ can be deduced. Here comes the third, and hardest, step of the story: understanding the possible \emph{mass deformations} of the scale invariant geometries. As we said above, turning on ($\cN=2$ SUSY-preserving) relevant operators does not lift the CB, and it instead deforms it in a precise manner.  The different possible deformations thus give different continuous families of CB geometries (depending on the mass parameters) which can lift the degeneracy between distinct SCFTs which have the same scale-invariant CB geometry. In the rank-1 case, this deformation information can be encoded via the \emph{deformation pattern} of the scale invariant geometry. If refined with the deformation pattern, the initial scale invariant CB geometry data, almost uniquely characterizes a $\cN=2$ SCFT.\footnote{It is known that this data is not enough to distinguish also discretely gauged theories.} 

In summary, our classification program then consists in picking an increasing CB complex dimension, determining the allowed scale invariant geometries, and then understanding their possible mass deformations.


\begin{table}[ht]
\centering \small
$\def\arraystretch{1.0}
\begin{array}{clcc|c|cc:c|ccc|}
&\multicolumn{3}{l|}{\text{CB:}} &
\multicolumn{1}{l|}{\text{HB:}} &
\multicolumn{3}{l}{\text{ECB \&\ flavor symm.:}} & 
\multicolumn{3}{l}{\text{Central charges:}} 
\\[1mm]
&\text{SI sing.} & \D(u) &\text{deform.} 
&\ \ d_{\text{HB}}\ \  
&\ \  h\ \ &\ \ {\bf 2h}\ \ &\quad \ff\quad 
&\ \ k_\ff\ \ &\ \ 24a\ \ &\ \ 12c\ \  
\\[1.5mm]
\hline\hline
&&&&&&&&&&\\[-4.5mm]
&II^* & 6 & \{{I_1}^{10}\}
& 29 & 0 & - & E_8 
& 12 & 95 & 62 \\
&III^* & 4 & \{{I_1}^9\}
& 17 & 0 & - & E_7  
& 8 & 59 & 38 \\
&IV^* & 3 & \{{I_1}^8\}
& 11 & 0 & - & E_6 
& 6 & 41 & 26 \\
\rcy &I_0^* & 2 & \{{I_1}^6\} 
& 5 & 0 & - & D_4 
& 4 & 23 & 14 \\
&IV & 3/2 & \{{I_1}^4\} 
& 2 & 0 & - & A_2
& 3 & 14 & 8 \\
&III & 4/3 & \{{I_1}^3\} 
& 1 & 0 & - & A_1 
& 8/3 & 11 & 6 \\
&II & 6/5 & \{{I_1}^3\} 
& 0 & 0 & - & \varnothing 
& - & 43/5 & 22/5  \\
\rcr \multirow{-9}{4mm}{\begin{sideways}$I_1$ series\qquad \  \end{sideways}}
&I_1 & 1 & - 
& 0 & 0 & - & U_1 
& * & 6 & 3  \\[.5mm]
\hline\hline
&&&&&&&&&&\\[-4.5mm]
&II^* & 6 & \{{I_1}^6,I_4\} 
& 16 & 5 & \bf10 & C_5 
& 7 & 82 & 49 \\
&III^* & 4 & \{{I_1}^5,I_4\} 
& 8 & 3 & (\bf6,1) & C_3A_1 
& (5,8) & 50 & 29 \\
&IV^* & 3 & \{{I_1}^4,I_4\} 
& 4 & 2 & {\bf4}_0 & C_2U_1
& (4,?) & 34 & 19 \\
\rcy &\blue{I_0^*} & 2 & \{{I_1}^2,I_4\} 
& 0 & 1 & \bf 2 & C_1 
& 3 & 18 & 9 \\
\rcr \multirow{-6}{4mm}{\begin{sideways}$I_4$ series\qquad \ \end{sideways}}
&I_4 & 1 & - 
& 0 & 0 & - & U_1 
& * & 6 & 3 \\[.5mm]
\hline\hline
&&&&&&&&&&\\[-4.5mm]
&II^* & 6 & \{{I_1}^3,I^*_1\} 
& 9 & 4 & \bf 4 \oplus \bar 4 & A_3{\rtimes}\Z_2 
& 14 & 75 & 42 \\
&III^* & 4 & \{{I_1}^2,I^*_1\} 
& ? & 2 & \bf 2_+ {\oplus}\, 2_- & A_1U_1{\rtimes}\Z_2 
& (10,?) & 45 & 24 \\
&\green{IV^*} & 3 & \{I_1,I^*_1\} 
& 0 & 1 & \bf 1_+ {\oplus}\, 1_- & U_1 
& * & 30 & 15 \\
\rcr \multirow{-5}{4mm}{\begin{sideways}$I_1^*$ series\qquad\  \end{sideways}} 
&I_1^* & 2 & - 
& 0 & 0 & - & \varnothing 
& - & 17 & 8 \\[.5mm]
\hline\hline
&&&&&&&&&&\\[-4.5mm]
&II^* & 6 & \{{I_1}^2,IV^*_{Q=1}\} 
& ? & 3 & \bf 3 \oplus \bar 3 & A_2{\rtimes}\Z_2 
& 14 & 71 & 38 \\
&\green{III^*} & 4 & \{I_1,IV^*_{Q=1}\} 
& 0 & 1 & \bf 1_+ {\oplus}\, 1_- & U_1{\rtimes}\Z_2 
& * & 42 & 21 \\
\multirow{-4}{4mm}{\begin{sideways}$\scriptstyle{IV^*_{\scriptscriptstyle Q=1}}$ \small{ser.}\quad \  \end{sideways}} 
&IV^*_{Q=1} & 3 & - 
& 0 & 0 & - & \varnothing 
& - & 55/2 & 25/2 \\[.5mm]
\hline\hline
\rcy &&&&&&&&&&\\[-4.5mm]
\rcy &\blue{I_0^*} & 2 & \{{I_2}^3\} 
& 0 & 1 & \bf 2 & C_1 
& 3 & 18 & 9 \\
\rcr \multirow{-3}{4mm}{\begin{sideways}$I_2$ ser.\quad\ \  \end{sideways}} 
&I_2 & 1 & - 
& 0 & 0 & - & U_1 
& * & 6 & 3 \\[.5mm]
\end{array}$
\caption{Partial list of rank-1 $\cN=2$ SCFTs.  They are divided into 5 series; the CFTs within each series are connected by RG flows from top to bottom.  The red rows give the characteristic IR-free theory each series flows to.  Yellow rows are lagrangian CFTs, while blue and green singularities have enhanced $\cN=4$ and $\cN=3$ supersymmetry, respectively.  The first 3 columns describe the CB geometry; the next column gives the Higgs branch dimension; the next 3 columns give properties of the mixed branch and the flavor symmetry; and the last 3 columns give the CFT central charges.  The meaning of each column and the choice of the theories appearing in the rows are explained in the introduction to \cite{Argyres:2016xmc}.\label{tab1}}
\end{table}

\paragraph{Rank-0 theories.}  

An $\cN=2$ theory with no CB is a rank-0 theory and there is a belief that no interacting rank-0 $\cN=2$ SCFT exists. This is largely based on the lack of counter-examples and could be a consequence of a \textit{lamp post effect}; many techniques to study $\cN=2$ SCFTs are based on assuming the existence of the CB. There is also some further evidence from our rank-1 classification (described more below). In carrying out this classification we explicitly assume no interacting rank-0 SCFTs exist. If that were not the case, and a rank-0 SCFT with small enough central charges did instead exist, our results would be modified in a dramatic way.\footnote{For more details on this point see the discussion in section 5 of \cite{Argyres:2015gha}.} Instead of predicting the existence of a total of 28 theories,\footnote{The exact number of rank-1 theories has been the source of some confusion, particularly since each of the summary tables in \cite{Argyres:2015ffa,Argyres:2015gha,Argyres:2016xua,Argyres:2016yzz,Argyres:2016xmc} seem to report contradicting results. 28 is the number of Coulomb branch geometries of non-discretely gauged theories which certainly exist.} the number would be close to a hundred and those new rank-1 theories would have a CB and therefore would be detectable with a multitude of methods. The fact that our classification appears to be complete and that there is no sign of the existence of these extra theories, therefore provides evidence supporting the \textit{non-existence of rank-0} theories conjecture, at least within a certain range of central charges. By extending our systematic classification to higher ranks, we can considerably strengthen this indirect evidence. 



 
\section{Review of important results}

\begin{wrapfigure}{L}{0.50\textwidth} 
\includegraphics[width=.35\textwidth]{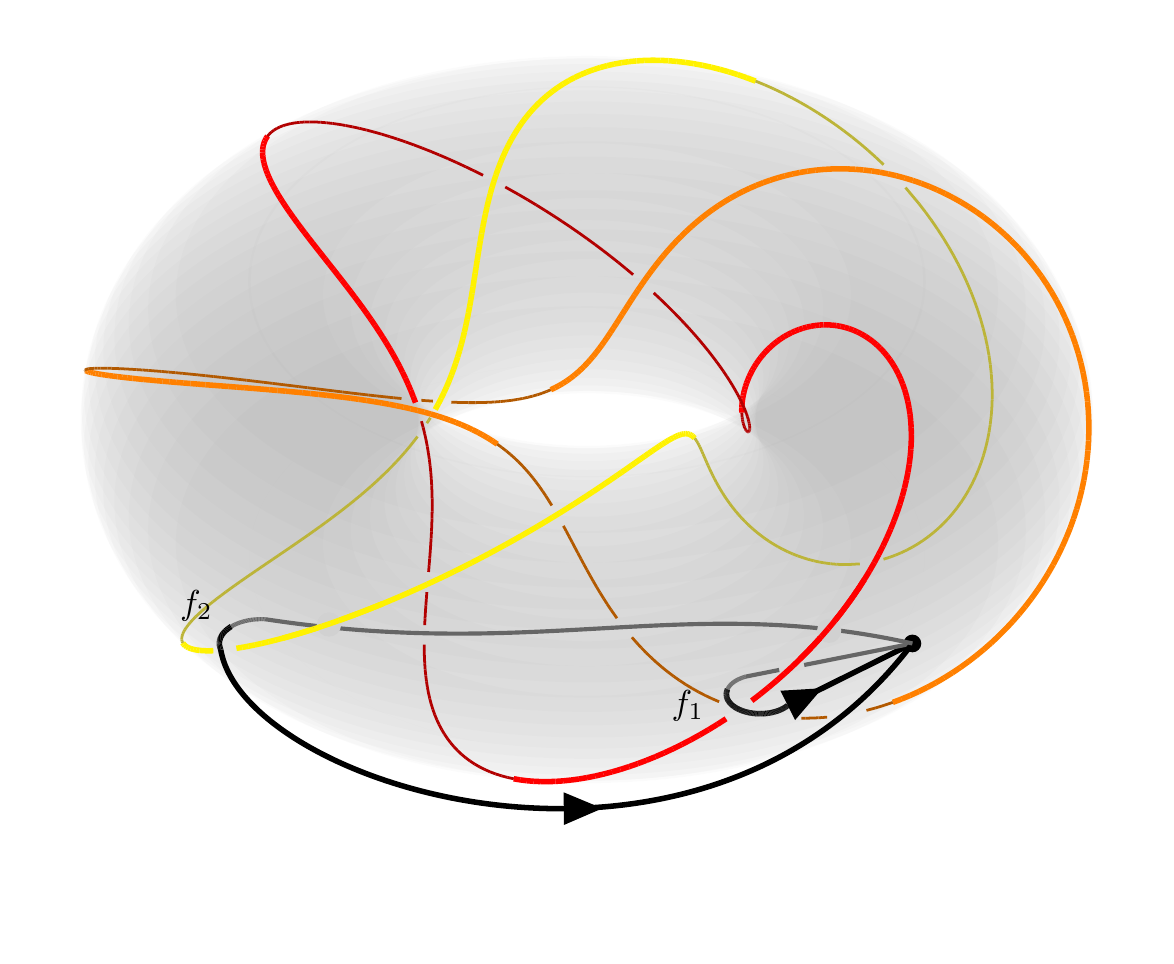}
\caption{Depiction of an $L_{(1,2)}(0,3,0)$ torus link consisting of the red, orange, and yellow circles.}
\label{knot3}
\end{wrapfigure}

\paragraph{Full classification of rank-1 geometries.} 

In a series of papers \cite{Argyres:2015ffa,Argyres:2015gha,Argyres:2016xua,Argyres:2016xmc} the program outlined above was carried out completely in the one complex dimensional CB case which led to a complete classification of rank-1 theories. The results of our analysis are summarized in table 1 of \cite{Argyres:2016xmc} which is reported here, see table \ref{tab1}. As discussed extensively in \cite{Argyres:2016yzz}, it is possible to start from some of the theories in table \ref{tab1} and gauge discrete subgroups without breaking $\cN=2$ SUSY. This operation acts non-trivially on the CB but without lifting it, so it produces other rank-1 theories. This explains why table 1 in \cite{Argyres:2016yzz} differs from table \ref{tab1} here. Recently it was shown that all the rank-1 theories can be obtained from 6 dimension. In particular, starting from specific 6d (1,0) theories compactified on a $T^2$ and twisted by non commuting (flavor) holonomies, it is possible to obtain the theories sitting at the top of each one of the series in table \ref{tab1} \cite{Ohmori:2018ona}. The rest can be obtained by turning on specific subsets of their mass deformations. \cite{Apruzzi:2020pmv} discusses instead the construction of all entries in table \ref{tab1} in F-theory. To the best of our knowledge, all the known rank-1 $\cN=2$ SCFTs are captured by our analysis.

\paragraph{Understanding the singularity structure of rank-2 geometries.} 

The analysis of scale invariant CB geometries at complex dimension two is already considerably more challenging than the rank-1 case outlined above. In \cite{Argyres:2018zay} we were able to show that the $\C^*$ action, along with a basic assumption to avoid pathological behavior of the CB (such as singularities dense in $\cC$),\footnote{Though we feel strongly that such behaviors are unphysical, we have thus far been unable to prove it \cite{Argyres:2018zay}.} constrains dramatically the topology of $\cV_{\rm rank-2}$. Specifically we showed that $\cV_{\rm rank-2}\cap S^3$, where the intersection with the three sphere $S^3$ is taken to get rid of the contractible direction corresponding to the scaling action, is in general a $(p,q)$ torus \text{$n$-link}, perhaps with additional unknots \cite{Argyres:2018zay}; see fig \ref{knot3}.

The set of allowed $(p,q)$ is strongly restricted by the $\Sp(4,\Z)$ monodromy structure and only a finite set of values is allowed. Yet the number, $n$, of components of the link is not constrained in any obvious way by the single monodromies. In particular $n$ could be infinite therefore leading to an infinite set of topologically inequivalent CB geometries. A possible restriction could come from studying the whole monodromy group, and not just individual monodromy elements. The former provide a representation of the fundamental group of the smooth part of the CB $\cC\setminus \cV$ and thus is sensitive to global data. The fundamental group of the spaces of interest for rank-2 was computed recently \cite{Argyres:2019kpy}.

\paragraph{Metric vs. complex singularities.}  

The singularities on $\cV$ can occur in two types: \textit{metric} singularities, $\cV_{\rm metric}$, and singularities in the \textit{complex structure}, $\cV_{\rm cplx}$. At $\cV_{\rm metric}$, the CB as an algebraic (projective) variety is perfectly fine though the metric structure is non-analytic.\footnote{The CB is still a metric space at these points, but does not have a well-defined Riemannian metric.  Think of the tip of a 2-dimensional cone as an example.}  $\cV_{\rm cplx}$ is instead the set of points in which the CB is singular as an algebraic variety. The latter type were once believed not to occur but counter-examples were pointed out in \cite{Bourget:2019phe,Argyres:2018wxu}.

The physical interpretation of $\cV_{\rm metric}$ and $\cV_{\rm cplx}$ is considerably different. $\cV_{\rm metric}$ occur where charged states become massless.  This can only happen when the BPS lower bound on their mass vanishes which restricts $\cV_{\rm metric}$ to be complex co-dimension one in $\cC$. 

The locus of complex singularities $\cV_{\rm cplx}$ is generically a proper subvariety of the locus of metric singularities.  $\cV_{\rm cplx}$ occur when the Coulomb branch chiral ring of the $\cN=2$ SCFT is \emph{not freely generated}. In this case new phenomena can take place like an apparent violation of the unitarity bound \cite{Argyres:2017tmj}. All the known cases of non-freely generated chiral rings arise by a non-trivial action of a discrete group on the CB \cite{Argyres:2019ngz,Bourget:2019phe,Argyres:2018wxu}. 

\paragraph{Scaling dimensions of Coulomb branch coordinates.} 

If the CB chiral ring is freely generated ($\cV_{\rm cplx}=\varnothing$), the scaling dimensions, $\D_i$, of the CB coordinates are proportional to the $\U(1)_r$, $r_i$, charges of the CB operators of the $\cN=2$ SCFT: $\D_i\propto r_i$. Superconformal representation theory only constrains  these charges to be greater than one, $\D_i\ge1$.  A relatively elementary argument shows that these scaling dimensions are instead constrained by the low energy electromagnetic monodromy group and that they belong to a finite set of rational numbers, whose size depends on the rank $r$ of the theory. Explicitly, the allowed values for the $\D_i$ of a rank-$r$ theory are \cite{Argyres:2018urp,Caorsi:2018zsq}:
\begin{align}
\D \in \left\{ \frac{n}{m} \ \bigg\vert\ n,m\in\N,\ 0<m\le n,\ 
\gcd(n,m)=1,\ \vf(n) \le 2r \right\}
\nonumber
\end{align}
where $\varphi(n)$ is the Euler totient function and the maximal dimension allowed grows superlinearly with rank as $\D_\text{max} \sim r \ln\ln r$.  

If the CB chiral ring is not freely generated, the relation between scaling dimension of CB coordinates and $\U(1)_r$ charges of CB multiplet is less straightforward. Often (maybe always) the scaling dimensions are neither globally nor uniquely defined.

\section{Future developments}

\paragraph{Finite vs.\ infinite number of allowed geometries.} 

One of the motivating reasons behind carrying out the program of classifying $\cN=2$ SCFTs in four dimensions is the belief that at any given rank only a finite set of such theories exist.\footnote{We are counting families of SCFTs connected by exactly marginal deformations as a single SCFT for the purpose of this counting.}  If this belief is correct, it trivially follows that only a finite set of scale invariant CB geometries are \emph{realized} as moduli spaces of consistent physical theories at any given rank. The surprising fact that all the CB geometries allowed in rank-1 are indeed realized, motivates instead the authors' belief that in fact only a finite number of scale invariant CB geometries are \emph{allowed}. As mentioned above, our analysis of rank-2 scale invariant CB geometries, arrives close to showing that this is the case for two complex dimensional CBs, but we fall short of showing that for any given value of $(p,q)$ only a finite number of link components is allowed. If that were achieved it would be an important conceptual result.

It is important to also stress, that proving the existence of a finite set of scale invariant CB geometries at any given rank, is only a necessary condition for showing that the at that rank only a finite number of $\cN=2$ SCFTs exist. In fact it is logically possible, that a given scale invariant geometry admits an infinite set of inequivalent mass deformations and therefore corresponds to an infinite set of physically distinct $\cN=2$ SCFTs.

\paragraph{Stratification of the Coulomb branch and importance of rank-1 theories.} 

We have thus far said little about the structure of the singular locus $\cV$. If we assume, again, that some pathological behaviors are avoided and in particular that $\cV$ is a complex co-dimension one complex subvariety of $\cC$,\footnote{$\cV$ is identified by the zeros of the central charge $Z_Q$, where $Q$ is the electromagnetic charge of a populated BPS state. If $Z_Q$ were a function on $\cC$ then it would be easy to prove that $\cV$ is indeed a complex subvariety. But $Z_Q$ is indeed branched over $\cV$ and thus the challenge in showing this result in generality.} then it is possible to convincingly argue that $\cV$ is itself an RSK variety of one complex dimension less, or more specifically the union of a finite number of such RSK varieties. This result might sound counter-intuitive. But the observation that there cannot be a transition among two inequivalent vacua at zero energy cost, implies that it has to be possible to induce a non-zero metric on $\cV$ from the ambient space $\cC$. This can be done by identifying an appropriate set of complex coordinates on $\cC$, ($u^\perp,u^\parallel)$, such that $\cV$ is at $u^\perp=0$. The metric is induced by considering the $\partial_{\parallel}\sigma$ and $\partial_{\parallel}\bar{\s}$ components, which are well-defined on $\cV$, where $\s$ labels the vector of special coordinates. 

It is also possible to show that the entire RSK structure can be restricted to $\cV$, thus providing a consistent lower-dimensional RSK space. For more details on the argument see \cite{Argyres:2018zay} and in particular \cite{Argyres:2019yyb} where the stratification of the singular locus is discussed in the context of $\cN=3$ theories. This discussion, with some minor but important modifications, carries over to the $\cN=2$ case. This powerful result, which parallels the structure of singularities of symplectic varieties, opens exciting perspectives on carrying out the classification of higher-dimensional geometries using a sort of inductive argument starting from the completed rank-1 story.

\paragraph{Polarization and other $\cN=4$ theory.} 

The RSK structure on $\cC$ is richer than we have thus far discussed. In fact the $U(1)^r$ low-energy physics provides a natural integral skew-symmetric pairing on the special coordinates. This arises as follows. States in the low-energy theory are labeled by a set of $2r$ integers, their corresponding electric and magnetic charges which we will collectively label $Q$.\footnote{Recall that on $\cC_{\rm reg}$ these states are massive; the theory there has no massless charged state.}  Denote by $\langle Q, Q'\rangle:=Q^T\mathbb{D} Q'$, where $\mathbb{D}$ is an integer non-degenerate skew-symmetric $2r\times 2r$ matrix in canonical form, the pairing induced by the Dirac-Zwanziger-Schwinger quantization condition on the lattice of electric-magnetic charges. Since the special coordinates are dual to the lattice of electric-magnetic charges, $\mathbb{D}$ induces a pairing on them as well. We call $\mathbb{D}$ the \emph{polarization} of the lattice of electric-magnetic charges, and it has the very important property of determining the structure of the electric-magnetic duality group, which is indeed $\Sp_\mathbb{D}(2r,\Z)$.

If the polarization can be brought to the canonical form $\mathbb{D
}=\epsilon\otimes \I_r$, where $\epsilon$ is a $2\times2$ antisymmetric matrix, $\mathbb{D}$ is called \emph{principal} and $\Sp_\mathbb{D}(2r,\Z)$ reduces to the more standard $\Sp(2r,\Z)$. Theories with non-principal polarization have not been studied in any real detail. There is some mild evidence, from the classification of rank-1 geometries, that such theories are indeed relative field theories \cite{Argyres:2015ffa,Caorsi:2019vex}. This result is very preliminary and this exciting subject certainly deserves more study.

\subsubsection*{Acknowledgements}

We would like to thank our collaborators (listed in the references) as well as O. Aharony, C, Beem, S. Cecotti, M. Caorsi, M. Del Zotto, J. Distler, I. Garcia-Extebarria, S. Giacomelli, A. Hanany, M. Lemos, C. Meneghelli, L. Rastelli, S. Schafer-Nameki, Y. Tachikawa, and T. Weigand for many helpful discussions and insightful comments. This work benefited from the 2019 Pollica summer workshop, which was supported in part by the Simons Foundation (Simons Collaboration on the Non-perturbative Bootstrap) and in part by the INFN. The authors are grateful for this support.  PCA is supported by DOE grant DE-SC0011784, and MM is  supported by NSF grants PHY-1151392 and PHY-1620610.

\footnotesize
\bibliographystyle{bib}
t
\end{document}